\documentclass[aps,twocolumn,pra,10pt,nofootinbib,floatfix,superscriptaddress]{revtex4-2}

\usepackage[T1]{fontenc}
\usepackage{libertinus}
\usepackage{libertinust1math}

\usepackage[a4paper,left=1.8cm,right=1.8cm,top=1.8cm,bottom=2cm]{geometry}

\usepackage{graphicx}
\usepackage{amsmath}
\usepackage{amssymb}
\usepackage{float}   
\usepackage{wrapfig}
\usepackage{tcolorbox}

\usepackage[normalem]{ulem}
\usepackage{titlesec}
\titlespacing{\section}{0ex}{0ex}{1ex}
\titlespacing{\subsection}{0ex}{2ex}{1ex}

\titleformat{name=\section}{\large \bfseries}{\thesection}{1ex}{}
\titleformat{name=\subsection}{\bfseries}{\thesubsection.}{1ex}{}

\usepackage{xcolor}
\usepackage[colorlinks = true, citecolor={blue!45!red}, linkcolor={blue!45!green}, urlcolor={blue!65!black}, breaklinks=true, linktocpage=true]{hyperref}
\usepackage{cleveref}

\newfloat{extdata}{htbp}{loe}
\floatname{extdata}{Supplementary Fig.}
\crefname{extdata}{Extended Data Fig.}{Extended Data Figs.}

\newcommand{\mLLG}{nM-LLG}
\newcommand{\meth}{Methods}
\newcommand{\supp}{Appendix}
\newcommand{\alp}{{\cal A}}
\newcommand{\EDF}{Appendix Fig.}

\begin{document}

\title{Intrinsic non-Markovian magnetisation dynamics}

\author{F. Hartmann\normalfont\textsuperscript{$\star$}}
\email{hartmann3@uni-potsdam.de}
\affiliation{Institute of Physics and Astronomy, University of Potsdam, 14476 Potsdam, Germany}

\author{V. Unikandanunni\normalfont\textsuperscript{$\star$}}
\affiliation{Department of Physics, Stockholm University, Stockholm, Sweden}
\affiliation{Institute of Applied Physics, University of Bern, Sidlerstrasse 5, 3012 Bern, Switzerland}

\author{M. Bargheer}
\affiliation{Institute of Physics and Astronomy, University of Potsdam, 14476 Potsdam, Germany}
\affiliation{Helmholtz-Zentrum Berlin für Materialien und Energie GmbH, Wilhelm-Conrad-Röntgen Campus,
BESSY II, 12489 Berlin, Germany}

\author{E. E. Fullerton}
\affiliation{Center for Memory and Recording Research, University of California San Diego, San Diego, California 92093, USA}

\author{S. Bonetti}
\affiliation{Department of Physics, Stockholm University, Stockholm, Sweden}
\affiliation{Department of Molecular Sciences and Nanosystems, Ca’ Foscari University of Venice, 30172 Venice, Italy}

\author{J. Anders}
\email{janet.anders@uni-potsdam.de}
\affiliation{Institute of Physics and Astronomy, University of Potsdam, 14476 Potsdam, Germany}
\affiliation{Department of Physics and Astronomy, University of Exeter, Exeter EX4 4QL, UK}

\def\thefootnote{$\star$}\footnotetext{These authors contributed equally to this work.}

\begin{abstract}

\noindent Memory effects are widespread in complex systems while evidence of non-Markovian dynamics in fundamental physical systems has been limited to specifically engineered cases. Here we report on the experimental observation of non-Markovian magnetisation dynamics in crystalline cobalt. Intense terahertz electromagnetic driving brings the system into a non-equilibrium state, and allows us to follow its magnetisation dynamics in real time at low temperatures. The response reveals a striking multi-peaked frequency spectrum that defies conventional descriptions. 
Using open quantum system theory we evidence that a non-Markovian memory kernel, arising from spin–phonon bath coupling, naturally explains both the spectral structure and its temperature dependence. 
Theoretical prediction of the spectrum is accomplished by   simulating the non-Markovian Landau-Lifshitz-Gilbert equation, while analytics quantitatively determines the frequencies of three spectral peaks.
The results demonstrate that intrinsic memory effects can emerge in a simple ferromagnet, opening new opportunities to explore and control non-Markovian dynamics in condensed-matter systems.

\end{abstract}

\maketitle

\noindent

\begin{figure}[b!]
    \centering \includegraphics[width=0.44\textwidth]{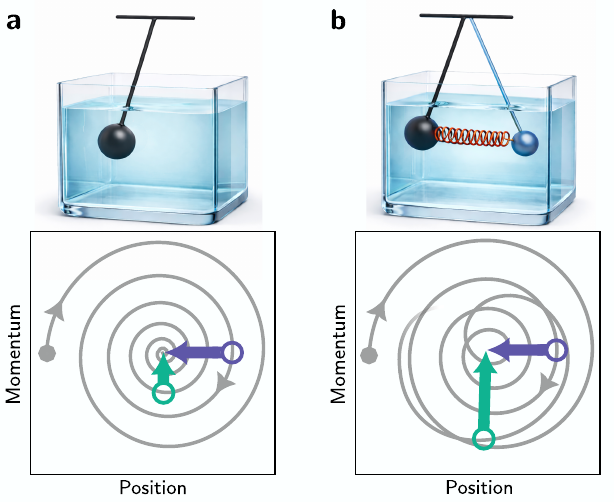}
    \caption{\textbf{Sketch of Markovian vs non-Markovian dynamics.} \textbf{a.} A pendulum (black) interacting with its environment (illustrated as blue fluid) has a trajectory (grey spiral) in position-momentum space that monotonously moves towards its state of rest, consistent with Markovian dynamics. 
    {\textbf b.} In contrast, the trajectory (grey spiral-like) of a single pendulum (black) that interacts (orange spring) with another pendulum, shows complex dynamics with non-Markovian features, such as time excursions away from its state of rest. This is highlighted by the later-time arrow (green) which is longer than the earlier-time arrow (purple), each showing the distance from the state of rest.}
    \label{fig:nonMarkovianHO}
\end{figure}

\noindent Damping is an everyday phenomenon, which we experience when we need to pedal to keep a constant speed on a bicycle, or in the movement of an empty swing that climbs less high with every round, finally coming to rest.
Microscopically, damping arises from the system continuously interacting with a large number of  degrees of freedom, its environment or bath. 
Such dissipative dynamics is usually one-way: the system will lose energy to the bath and move closer to the rest state,
as illustrated for a single pendulum in Fig.~\ref{fig:nonMarkovianHO}{\textbf a}. 
This is known as Markovian dynamics where the bath is memoryless.
In contrast, when two individual pendula are coupled to each other, they are able to transfer energy back and forth. Each pendulum then undergoes so-called ``non-Markovian'' dynamics, where backflow of energy is possible. Much more complex motion then emerges, sometimes towards and sometimes away from the rest state, see Fig.~\ref{fig:nonMarkovianHO}{\textbf b}.
Indeed, such memory-effects arise in many complex systems that have interacting components, including the spreading of epidemics in a population~\cite{Sofonea_2021} and financial decision making~\cite{Wachter_2023}.
However, non-Markovian dynamics can in principle also arise in a fundamental physical system through the coupling to its dissipative bath.

This requires the bath to play a much more active role; instead of just receiving dissipated energy, it must be able to give energy back. In statistical physics, such dynamics is modeled by a generalised Langevin equation~\cite{Loos2025}, which includes a memory kernel $K(t-t')$. This function weights the impact of the past state of the system (at time $t'$) on the current dynamics (at time $t$). 
For kernels with short memory times, the dynamics reduces to the standard (Markovian) Langevin equation.
%
%
For example, protein folding dynamics data has been shown to be matched by models with a non-trivial kernel~\cite{Ayaz_2021}, and methods to accurately infer memory kernels from observed system dynamics have been developed~\cite{Tepper_2024}. 
Meanwhile, most experiments demonstrating  non-Markovianity have been confined to situations akin to the two pendula in Fig.~\ref{fig:nonMarkovianHO}{\textbf b}. For example, for a Nitrogen-Vacancy (NV), the dynamics of its electron spin is non-Markovian due to its tunable interaction with the NV centre's nuclear spin~\cite{Haase_2018}, while any remaining bath-influence is  demonstrably Markovian.
Experimental evidence of non-Markovian dynamics that arises purely from a system's interaction with a dissipative bath, has been confined to  rather complex and often highly engineered systems~\cite{Madsen_2011,Liu_2011,Gröblacher_2015,Potonik_2018,Odeh_2025,Ginot_2025}, such as quantum dots in microfabricated nanostructures~\cite{Madsen_2011}, 
photon polarisation made to interact with photon frequency in all-optical setups~\cite{Liu_2011}, and colloidal particles swimming in polymer enriched viscoelastic baths~\cite{Ginot_2025}.

Here we report on the discovery of intrinsic non-Markovian signatures in a fundamental condensed matter system: a ferromagnetic cobalt film. 
The quantity of interest is the magnetisation $\mathbf{M}$ of the sample, while the bath that damps its dynamics are the natural vibrational excitations of the solid (i.e. phonon bath)~\cite{Strungaru_2021}.
By driving with an intense terahertz pulse, we bring the spin system out of equilibrium and approximately into resonance with the phonon bath with which it can exchange energy. This allows the bath to not only damp, but also return energy and information back to the system.
Our experimental magnetisation signal shows a complex Fourier spectrum with multiple peaks, which are unexplained by established theory, such as the Landau-Lifshitz Gilbert (LLG) equation~\cite{Lakshmanan_2011} and the inertial LLG (iLLG) equation~\cite{Ciornei_2011,Li_2015, Neeraj_2020,Unikandanunni_2022,Mondal_2023,De_2025}.
To explain the peaks, we advance that the missing conceptual ingredient is the influential role of the bath of lattice vibrations.
Taking an open quantum system approach widely used in quantum optics~\cite{barnett1997methods,Huttner_1992}, we model the spin interaction with its phononic environment~\cite{Anders_2022}.
Parametrising the resulting non-Markovian LLG (\mLLG) equation with experimental values, we find that the simulated spectrum shows excellent agreement with the experimental spectrum.

\medskip 

\section*{Results}

\noindent {\bf Pump-probe experiment}

\noindent We perform single-cycle THz pump/time-resolved magneto-optical Kerr effect (MOKE) probe measurements on fcc Co thin film samples at cryogenic temperatures (see~\meth). The film is epitaxially grown on an MgO substrate and exhibits exceptional crystalline quality (see \meth). 
The sample lies in the $x$-$y$-plane and is placed in a low-vibration cryostat with windows transparent to both the terahertz pump and the near-infrared probe light.
Single-cycle THz pulses are generated using organic crystals that provide highest spectral power density in the 1–3 THz range~\cite{hauri2011strong,jazbinsek2019organic}.
The THz magnetic field points along the $y$-direction, while the external bias field acts in the $x$-direction (see Fig.~\ref{fig:experimental_setup} in~\meth). The THz pulse resonantly excites spin dynamics while minimizing thermal effects.

\begin{figure}[t]
\centering \includegraphics[width=0.46\textwidth]{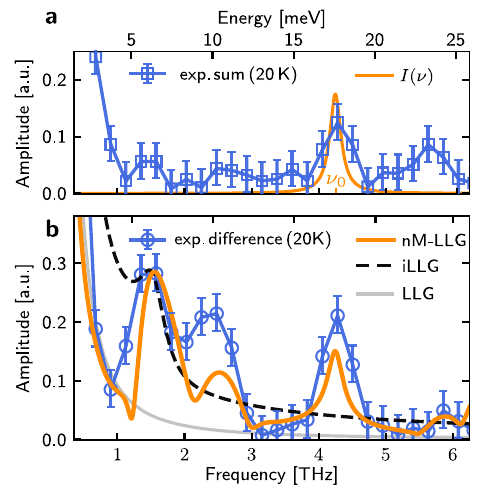}
    \caption{{\bf Sum and difference spectra of terahertz-induced dynamics in a ferromagnetic cobalt film.} 
   \textbf{a.} Fast Fourier transform of summed (non-magnetic) time-resolved signal (blue symbols on  solid line)  following spin excitation with a strong terahertz pump field. 
    Lorentzian spectral density $I(\nu)$ (orange solid line) centred at frequency $\nu_0$  = $4.2$~THz to match signal, and with width $\Gamma = 0.2$~THz, see main text.
\textbf{b.} Fast Fourier transform of the difference (magnetic) response following the same pump field and at the same temperature (blue symbols on solid line),  Fourier spectra of the simulated LLG~equation (grey solid line), iLLG~equation (black dashed line) and of the \mLLG~equation (orange solid line).  \mLLG~simulation solves  Eq.~\eqref{eq:mLLG} with memory kernel $K(t-t')$ specified by Lorentzian spectral density $I(\nu)$, with fixed parameters $\nu_0$, $\alp$ and $\Gamma$, see main text. Units for amplitudes of the experimental and simulation spectra are chosen such that the peak amplitudes at $1.4$~THz coincide. 
Whiskers indicate $2\sigma$ statistical error due to instrument noise. Additional leakage errors due to imperfect polarity flipping are not included.
}
    \label{fig:multipeak} 
\end{figure}

To isolate ``magnetic'' and ``non-magnetic'' parts of the signal at low temperatures, where cryostatic confinement makes switching the polarity of the bias magnetic dc field~\cite{Unikandanunni_2022,hudl2019nonlinear} impossible, we instead take two traces with flipped polarity of the terahertz pump field ($\pm \mu_0 \, H_{\sf THz} \, \mathbf{\hat{y}}$), see \EDF~\ref{fig:positive_and_negative_signal}.
By cross-checking the difference signal obtained using this method with that measured using the bias magnetic dc field switching method~\cite{Unikandanunni_2022}, both at room temperature, their equivalence is confirmed experimentally.
This difference (antisymmetric) signal isolates response components that are odd with respect to the THz field, which are primarily magnetic signatures. In contrast, the sum signal is dominated by non-magnetic contributions, such as phonon excitations. Thus, this approach provides selective access to signals primarily associated with magnetic (difference) and non-magnetic (sum) parts of the sample's response. Identifying the non-magnetic part, i.e. oscillatory features observed on sub-10-ps timescales, is important for parametrising the bath, as discussed below.

Typical time traces for the sum and difference signal, measured at temperature $T_{\sf exp}=20$~K, are shown in  \EDF~\ref{fig:sup3}. 
We take the fast Fourier transform (FFT) of each signal after the THz pulse has passed. Statistical errors due to instrument noise are propagated from time to Fourier domain, see \supp, and indicated by whiskers in Fig.~\ref{fig:multipeak}.
Additional leakage errors can occur between the magnetic and non-magnetic signals. These arise because when taking two THz field polarization traces, the experiment is not 100\% flipped (small change in THz polarization and THz generation efficiency) due to inhomogeneity in the organic crystal.

Fig.~\ref{fig:multipeak}{\bf a} shows the magnitude of the FFT corresponding to the non-magnetic contribution. The data reveal the presence of two peaks clearly outside the statistical error bars at 4.2~THz and 5.7~THz. This signal evidences phonon activity linked to the non-equilibrium magnetisation dynamics. The physical insight one gains is that the specific frequencies match the zone edge frequencies of the $\Gamma-L$ and $\Gamma-X$ phonons in fcc cobalt, respectively~\cite{Lizrraga_2017,Shokeen_2024} (see \supp). 
Identification of phonon activity at these frequencies is also consistent with prior \textit{ab initio} calculations~\cite{Carva_2013} based on Eliashberg theory, which show high electron-phonon coupling at these frequencies.

Fig.~\ref{fig:multipeak}{\bf b} shows the FFT of the magnetic part of the signal. It reveals a rich terahertz spectrum with three rather unexpected pronounced peaks, at ca. 1.4~THz, 2.4~THz and 4.2~THz.
Until recently, modelling of magnetisation signals has been accomplished with the LLG equation~\cite{Lakshmanan_2011}, a  phenomenological and Markovian theory describing precession and damping. However, it does not predict any such peaks.
Extending the LLG theory by including the angular momentum degree of freedom was proposed by Ciornei, Rub\'i and Wegrowe \cite{Ciornei_2011} resulting in the so-called inertial LLG equation, or iLLG.
This inertial theory~\cite{Ciornei_2011,Mondal_2023} predicts a single, much faster rocking motion in the THz range, on top of the precession typically observed in the GHz range in ferromagnets.
Such an effective spin inertia has been attributed to the coupling of spins to higher-order phonon-bath modes~\cite{Quarenta_2024}.
For a given inertial timescale $\tau_{\mathrm{in}}$ in the iLLG equation, the expected peak frequency in the magnetic response has been derived~\cite{Olive_2015,Cherkasskii_2020}.
In previous experiments, a single peak was observed experimentally following THz excitation~\cite{Neeraj_2020} as well as optical excitation of the electron system~\cite{De_2025}, consistent with the iLLG equation.
However, in addition to the inertial peak at about 1.4~THz the pronounced peaks that were observed in Ref.~\cite{Unikandanunni_2022}, and that we now observe at 2.4~THz and 4.2~THz, are not produced by either the LLG or iLLG~equations within physically reasonable parameter regimes (see \supp).

Recently a non-Markovian LLG~equation has been put forward, derived using an open quantum systems approach which models system and bath together~\cite{Anders_2022}.
In Fig.~\ref{fig:multipeak}{\bf b} we plot the Fourier spectrum of the magnetic response simulated with the \mLLG~equation as the orange solid line on top of the experimental data. It is immediately obvious that the \mLLG~simulation naturally produces an initial valley followed by multiple spectral peaks. Moreover, the quantitative match with the experimentally observed peak frequencies is outstanding. Simulation details are given in \meth. Simulations require three kernel parameters which are obtained from the experimental data, see below.
Additionally, analytic reasoning based on the \mLLG~equation provides a quantitative prediction of the three peak frequencies in the magnetic response.

\medskip 

\noindent {\bf Non-Markovian kernel}

\noindent The \mLLG~equation arises from coupling the spin system to the phonon bath~\cite{Anders_2022} via an interaction term
$V_{\sf int}~=~\mathbf{M}\cdot\int_0^\infty \mathrm{d}\nu \, \sqrt{\nu \, I(\nu)} \,\, \mathbf{X}_\nu$
where  $\mathbf{M}$ is the magnetisation of the sample,  $\mathbf{X}_\nu$ is the position coordinate of the phonon bath mode at frequency $\nu$, and $I(\nu)$ is the bath spectral density which characterises the strength of coupling of the spin system to various bath frequencies $\nu$. 
Solving the set of coupled equations for the magnetic system and the phonons results in the non-linear \mLLG~equation~\cite{Anders_2022} 
\begin{equation}
   \!\!\! \mathbf{\dot{M}}(t) = \mathbf{M}(t)\times\left[\gamma \mathbf{H}_{\sf eff}(t)  + \int_{t_0}^{t} \!\!\! \mathrm{d}t'K(t-t') \, \mathbf{M}(t') \right], 
    \label{eq:mLLG}
\end{equation}
where $\mathbf{M}(t)$ is the magnetisation at time $t$, $|\gamma|/2\pi = 28$~GHz/T,  $\mathbf{H}_{\sf eff}(t)$ is the effective magnetic field consisting of an external bias field  $\mathbf{H}_{\sf bias} = 0.1\,\mbox{T}\,\mathbf{\hat{x}}$ and a demagnetisation field $\mathbf{H}_{\sf d}(t)$ which acts substantially only along the $z$-direction (i.e. out-of-plane) in a thin film and is proportional to $M_z(t)$.

The non-Markovian nature of Eq.~\eqref{eq:mLLG} is made explicit by the presence of the memory kernel $K(t-t')$. Its significance is to link the dynamics at time $t$, to the magnetisation vector $\mathbf{M}(t')$ at previous times~$t'$. The strength and temporal extent of this memory is determined by the exact shape and parameters of the kernel. Physically, these parameters are fixed by the 
bath properties~\cite{Nemati_2022}. 
Knowing the bath spectral density $I(\nu)$
determines the memory kernel~\cite{Anders_2022}, $K(t-t') = 2\pi \int_0^{\infty} d \nu  \, I(\nu) \, \sin(2\pi \nu (t-t'))$ for $t \ge t'$. 
Spectral densities can be approximated with a Lorentzian shape~\cite{Anders_2022,Jankovi_2025} $I(\nu) = \alp\Gamma\nu/((\nu_0^2-\nu^2)^2+\Gamma^2\nu^2)$, which centers around a dominant  frequency $\nu_0$. $\Gamma$ and $\alp$ specify the range of frequencies where coupling is significant and the overall strength of the coupling, respectively. 

Zone-boundary and transverse acoustic phonons can carry energy and angular momentum in ultrafast demagnetisation experiments \cite{Dornes_2019,maldonado2020tracking,Shokeen_2024,Tauchert_2022}. Therefore, they can act as the energy and angular momentum bath in ultrafast spin dynamics.
Hence, in order to identify $I(\nu)$ for the sample measured, we return to the frequencies seen in the non-magnetic signal in Fig.~\ref{fig:multipeak}{\bf a}. 
The 4.2~THz peak is identified as the dominant bath contribution.
This peak corresponds to states occupied by zone-boundary transverse acoustic phonons~\cite{Lizrraga_2017}.
This naturally fixes the center frequency of the sample's Lorentzian spectral density $I(\nu)$ to $\nu_0 = 4.2$~THz. The remaining two Lorentzian parameters are obtained via known identities~\cite{Anders_2022}
$\alpha = \alp \Gamma / \nu_0^4$ and 
$\tau_{\mathrm{in}} = (\nu_0^2 - \Gamma^2) / (2\pi\nu_0^2 \Gamma)$. 
Here $\alpha$ and $\tau_{\mathrm{in}}$ are the damping parameter and the inertial time, 
previously found~\cite{Unikandanunni_2022} by fitting the first peak at $1.4$~THz in the experimental signal with the iLLG equation giving  $\alpha = 0.15$ and $\tau_{\mathrm{in}} = 0.8$~ps.
This fixes the kernel parameters to $\Gamma = 0.2$~THz and $\alp = 242.0$~THz$^3$. 
Numerical simulation of Eq.~\eqref{eq:mLLG}, including the memory kernel $K$ while not adding thermal noise (i.e. 0 K), produces the orange line in Fig.~\ref{fig:multipeak}{\bf b}. 
The match between fully parameterised simulation and experiment (at 20 K) provides compelling evidence that the spin system undergoes dynamics described by the non-Markovian LLG equation~\eqref{eq:mLLG}, which includes a temporally extended memory kernel. This non-Markovianity directly arises on the fundamental scale of the spins interacting with their phonon bath.

While the first peak is intuitively interpreted as the inertial signature, other peaks are not so readily given phenomenological interpretations. Indeed, further peaks arise  because equation~\eqref{eq:mLLG} is non-linear in spin as well as non-local in time, leading to highly complex dynamics.

\medskip 

\noindent{\bf Analytic reasoning for multiple peaks}

\noindent We now provide analytic reasoning for the appearance of multiple spectral peaks in Fig.~\ref{fig:multipeak}{\bf b}, followed by a quantitative explanation of the frequencies of the three peaks.

Returning to the \mLLG~equation~\eqref{eq:mLLG} it can be instructive to Taylor expand the magnetisation $\mathbf{M}(t')$ around time $t$ leading to multiple derivatives of $\mathbf{M}(t)$, i.e. 
\begin{equation}
    \mathbf{\dot{M}}(t) = \mathbf{M}(t)\times\left[\gamma\mathbf{H}_{\sf eff} (t) + \sum_{m = 1}^{\infty}  {\kappa_m} \, \partial_t^{m}\mathbf{M}(t) \right].
    \label{eq:mLLGexpanded}
\end{equation}
The derivatives are weighted by the coefficients $\kappa_m = (-1)^m  \int_0^{\infty}d \zeta \, \zeta^m \, K(\zeta)/ m!$, which are the $m$-th one-sided moments of the memory kernel \cite{Anders_2022}. 
The standard LLG equation \cite{Lakshmanan_2011} is captured by this description when only the first derivative (i.e. $m=1$) is included, which makes it Markovian by construction ~\cite{DeVega2017,Loos2025}. This approach also describes correctly the iLLG equation \cite{Ciornei_2011} which is the Taylor expansion up to the second derivative ($m=2$); see also~\cite{Mondal_2017} where inertial damping is identified as a higher order spin-orbit coupling effect.
Thus, the iLLG equation can be seen as a truncated Taylor expansion of the full non-Markovian description. 
We note that fractional derivatives (instead of the sum in Eq.~\eqref{eq:mLLGexpanded}) have recently been proposed~\cite{Verstraten_2023,Reyes-Osorio2025Bringing},  which also lead to non-Markovian dynamics.

For the iLLG equation, it has been argued~\cite{Cherkasskii_2020} that expressing the dynamics in the frequency domain, and by identifying the roots of the resulting frequency-polynomial, gives insight into where magnetic response peaks are expected. The LLG ($m=1$) and iLLG ($m=2$) equations produce one and two roots, respectively, which are commonly identified as the precession and the inertial peak. 
Applying this logic to the kernel expansion in \eqref{eq:mLLGexpanded} 
leads to the presence of higher derivatives which transform to higher order frequency polynomials in Fourier space. These have an increasing number of (often degenerate) roots.
%
%
Specifically, by making a sixth-order expansion in frequency $\nu$ of the Fourier transformation of a linear approximation to the temporal fluctuations of $\textbf{M}$ and $\textbf{H}$ in the non-linear \mLLG~equation, theory predicts peaks at 1.23~THz and 2.45~THz (see \supp).
Given the severity of the finite kernel expansion~\cite{Weissenhofer_2026} and its linearisation, we caution however, that the above polynomial cannot fully capture the mathematics of the exact non-Markovian kernel. 
A further magnetic response peak frequency is explained by observing that the full spin dynamics given by \eqref{eq:mLLG} contains an oscillating function, the kernel $K(t-t')$, which oscillates with frequency $\nu_1 = \sqrt{\nu^2_0-\Gamma^2/4} \approx$ 4.2~THz. The above three theoretically determined peak frequencies match well with the experimentally observed peaks at 1.4~THz, 2.4~THz, and 4.2~THz.

Kernel expansion \eqref{eq:mLLGexpanded} also provides an insightful analogy of spin dynamics with mechanics. While  acceleration $\ddot x$ is a standard quantity in the assessment of a particle's motion in one dimension $x$, time-derivatives higher than second order, such as $\overset{...}{x}$, $\overset{....}{x}$, are rarely discussed. Even their names, ``jolt'' and ``snap'', are hardly known. In  magnetization dynamics, the terms $\overset{...}{\mathbf M}, \overset{....}{\mathbf M}$, describe dynamical scales and retardation effects beyond the inertial term $\propto \overset{..}{\mathbf M}$, leading to increasingly complex motion. Just as in mechanics, these dynamics can be fully captured with numerics, while phenomenological approaches eventually buckle.

\begin{figure}
\centering \includegraphics[width=0.47\textwidth]{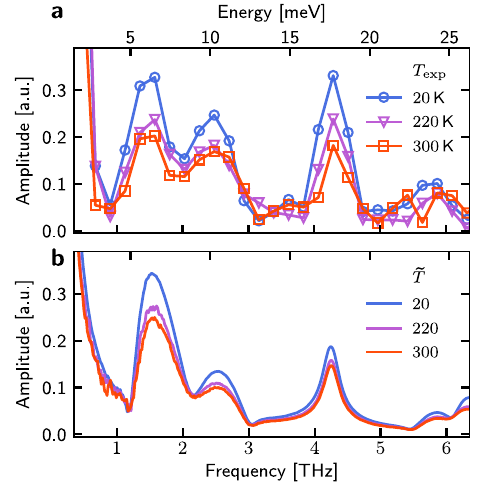}
      \caption{{\bf Temperature dependence of THz-frequency spectrum.} 
      \textbf{a.}  Fourier transform of experimental signal at three temperatures,  $T_{\sf exp} = 20$~K (blue line, circles), $T_{\sf exp} = 220$~K (purple line, triangles) and $T_{\sf exp} = 300$~K (red line, squares).  
      These three experimental spectra were measured with a single THz polarity, and contain magnetic and non-magnetic contributions.
      \textbf{b.} Fourier transform of dynamics obtained by solving the \mLLG~equation for three different unit-free temperatures $\tilde{T} \propto T_{\sf exp} $ chosen at the same ratios as in the experiment.   
      Simulations of the \mLLG~equation~\eqref{eq:mLLG} include coloured thermal noise, and the demagnetisation field is reduced for higher temperatures, see \supp. 
      Scale of the $\tilde{T} = 20$ graph in \textbf{b} is chosen such that first peak amplitude matches the experimental one in \textbf{a}. Amplitudes at higher temperatures follow relatively from it.
      All remaining simulation parameters are identical to Fig.~\ref{fig:multipeak}. 
      Instrument noise error bars (same as in Fig.~\ref{fig:multipeak}) are omitted for visual clarity of the data.}
    \label{fig:temperaturedep}
\end{figure}

\medskip 

\noindent {\bf Variation with temperature}

\noindent To complete the description of the bath-coupled dynamics, we finally explore the impact of temperature variation on the non-Markovian dynamics, i.e. variation of the population of vibrational modes in the bath. The experimentally measured spin dynamics at three temperatures, $T_{\sf exp} = 20$~K, $T_{\sf exp} = 220$~K, and $T_{\sf exp} = 300$~K, is presented  in the Fourier domain in Fig.~\ref{fig:temperaturedep}\textbf{a} (for the temporal traces see \EDF~\ref{fig:sup2}).
We observe that the frequencies of the spectral peaks remain fixed. Meanwhile, the peak amplitudes decrease when the temperature of the sample is increased. 

We now examine the expected temperature dependence arising from the \mLLG~equation~\eqref{eq:mLLG}. It incorporates thermal fluctuations via the inclusion of a thermal field $\mathbf{h}_{\mathrm{th}}(t)$ in $\mathbf{H}_{\sf eff}(t)$, which stochastically fluctuates in time~\cite{Anders_2022}. 
Additionally, the demagnetisation field is reduced at increased temperatures~\cite{Evans_2014}, see~\supp.

The noise $\mathbf{h}_{\mathrm{th}}(t)$ has, via fluctuation-dissipation theorem, the power spectral density $P(\nu) = 2 k_{\mathrm{B}}  T_{\sf exp} \, I(\nu)$, which introduces the experimental sample temperature $T_{\sf exp}$ while $k_{\mathrm{B}}$ is the Boltzmann constant.
Within the macrospin approximation,  precise information on the number of spins, their individual moments and their interactions is not available, which leaves the absolute temperature scale undetermined. Here we choose an absolute scale by setting the unit-free version of the simulation temperature, $\tilde{T}$, to $\tilde{T} = T_{\sf exp}/\mbox{K}$, which fixes the relative temperature ratios thereafter.
With temperature now included in the simulation of the \mLLG~equation~\eqref{eq:mLLG}, the THz-frequency spectrum for three temperatures, $\tilde{T} = 20, \,\tilde{T} = 220$ and  $\tilde{T} = 300$, is shown in Fig.~\ref{fig:temperaturedep}\textbf{b}.
As can be seen, the frequencies of the peaks remain fixed, while their relative amplitudes reduce with increasing temperature. The qualitative agreement with the experimental data in Fig.~\ref{fig:temperaturedep}\textbf{a} is also impressive.

\begin{figure}[t]
\centering \includegraphics[width=0.46\textwidth]{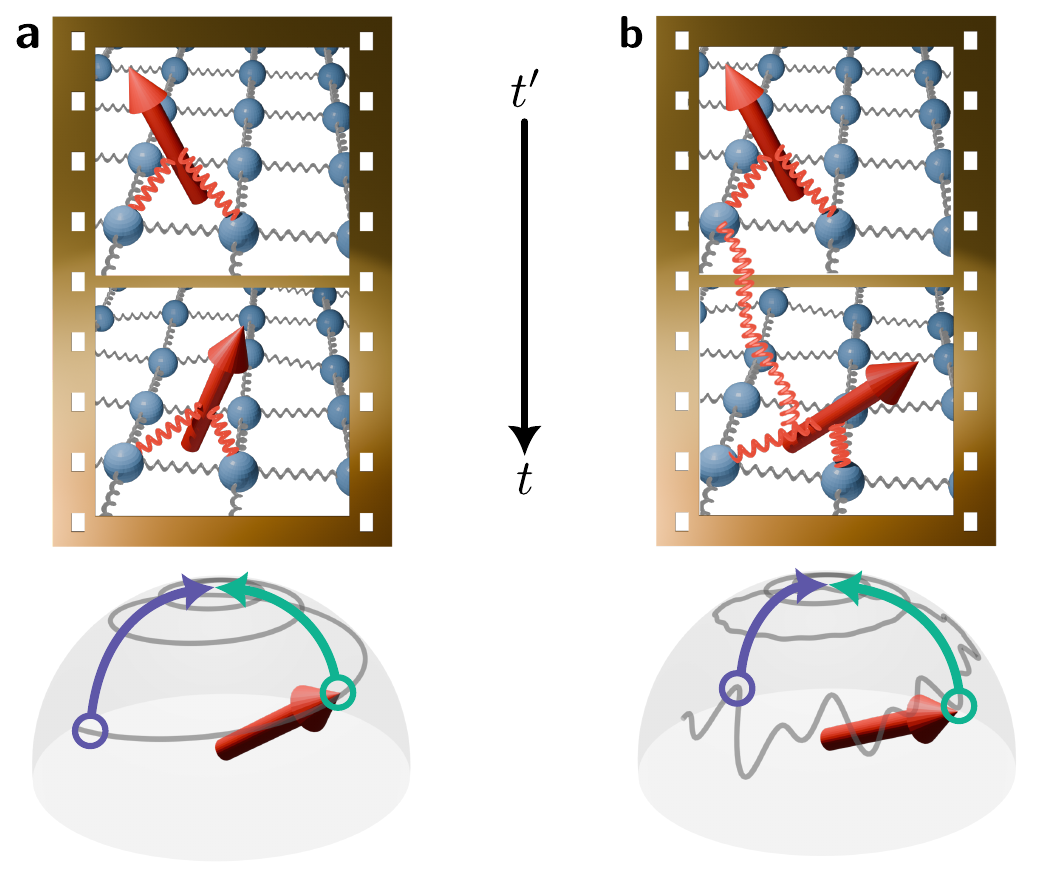}
      \caption{{\bf Illustration of Markovian and non-Markovian spin dynamics.} 
      \textbf{a}. Markovian dynamics: the magnetisation (red arrow) is coupled only to the instantaneous state of the phonon bath (connected grid) at each frame in time $t'$ and $t$. Viewed on the Bloch sphere, the magnetisation uniformly spirals towards its rest state (cf. Fig.~\ref{fig:nonMarkovianHO}\textbf{a}). 
      \textbf{b}. Non-Markovian dynamics: the magnetisation at a later time $t$ is coupled (red spring) also to the spin state at an earlier time $t'$. 
    Viewed on the Bloch sphere, the magnetisation dynamics exhibits a complex behavior
    (grey trajectory). The presence of  higher frequency dynamics leads to spin excursions away from its state of rest (green arrow), a clear non-Markovian characteristic. In the Fourier domain, such complex motion  presents in the form of multiple THz-frequency spectral peaks as seen in Fig.~\ref{fig:multipeak} and~\ref{fig:temperaturedep}. 
}
    \label{fig:spinphonon}
\end{figure}

\section*{Discussion}

\noindent The discovery that the collective dynamics of spins in a piece of ferromagnet is not described by the usual one-way damped motion is a major change of perspective for our understanding of condensed matter systems. In the new picture, the phonon bath plays a much more active role, not solely receiving dissipation from the spin system,  but also giving energy back. 
Such backflow occurs because the phonon bath, which interacts with the spin system,  imprints an intrinsic non-Markovian memory kernel on the non-linear spin dynamics. 
Theory also allows to estimate the kernel's memory time~\cite{Anders_2022}, i.e. the extent of a system's memory of its past, here $\tau_{\mathrm{memory}} = 2/\Gamma \approx 10$~ps. 
This natural picosecond memory time is much shorter than the time-period of precession in typical ferromagnetic resonance experiments,
which is in the nanosecond regime. 
On this slower timescale the coupling of the spins to the bath reduces to conventional damping, and non-Markovian effects cannot be observed, as illustrated in Fig.~\ref{fig:spinphonon}\textbf{a}.
In contrast, when non-equilibrium spin dynamics is excited in the THz range, as was done here, its timescale becomes of the same order as the bath's imprinted memory time.
Without time-scale separation, the past states of the spin  influence its future dynamics, as illustrated in Fig.~\ref{fig:spinphonon}\textbf{b}.
The measurable effect is a complex motion of the magnetisation at these time scales, which in the frequency domain shows up as multiple peaks in the THz regime.

We anticipate the presented findings to be of broader general validity, and to apply to the description of coupled collective dynamics in solids where measurements can be performed on the same timescale as the relevant memory kernel.
In the terahertz regime, this is expected to happen, for instance, in nonlinear phononics experiments, where phonons can exchange energy and angular momentum~\cite{minakova2025direct,Nova_2019,Mankowsky_2014}, while in  the gigahertz regime, non-trivial kernels could be present in coupled magnon-phonon dynamics~\cite{Berk_2019}.
Theoretical models predict non-Markovian signatures in the absorption spectra of monolayers of transition metal dichalcogenides, which feature strong exciton-phonon interaction~\cite{Lengers_2020}, and could be extended to insulating magnetic systems, such as YIG and GGG heterostructures, where anisotropic inertial effects arise~\cite{Quarenta_2024}.
Future experiments may also investigate predictions that non-Markovian baths can enable faster routes to stationarity~\cite{Strachan_2025}.
Considerable system-bath interactions promise a range of applications, such as improved control of system dynamics via the environment. Application areas include phonon-assisted coherent control of a single color center in van der Waals materials~\cite{Preuss_2022}, 
and reducing decoherence in quantum technologies with bath-engineering techniques~\cite{Odeh_2025}.
Non-Markovian baths are also expected to be important for control applications with atomic scale probes on surfaces~\cite{Chen_2022} as well as reducing environmental noise in quantum sensing with single spins~\cite{Zhou_2025,Rovny_2025,Chen_2022}.

By exciting and detecting dynamics on ultrashort timescales, our experiment opens new avenues for the exploration of regimes where the inanimate world of solids exhibits dynamics reminiscent of living systems, and where the monotonic growth of entropy no longer holds~\cite{Strasberg_2019Entropy,Skinner2021Improved,Hartmann_2026}. Identifying the right scales and mechanisms to observe such behaviour in other platforms will be a key next step. 

\medskip

\section*{Methods}
\noindent {\bf Experimental method}

\noindent We performed single-cycle THz pump-800~nm magneto-optical Kerr effect (MOKE) probe measurements on an epitaxially grown face-centered cubic cobalt sample. 
The sample was placed in a low-vibration cryostat from Advanced Research Systems, allowing temperature-dependent measurements. To minimize pump field loss, we installed TPX windows from \textit{Tydex} optimized for transmission at THz frequencies. The experimental schematic is shown in Fig.~\ref{fig:experimental_setup}. The magnetisation $\mathbf{M}$ of the sample is saturated along the $x$-direction using the field $\mu_0\mathbf{H}_{\mathrm{bias}}$ of $100$~mT of a permanent magnet mounted inside the cryo chamber on the sample holder. THz pump pulses required to drive the ultrafast magnetisation dynamics are generated in the organic crystal OH1 via nonlinear-optical rectification of $1300$~nm pulses from an optical parametric amplifier. The THz pulse's electric field is polarized along the $x$-direction with a maximum amplitude of 830 kV/cm, while the corresponding magnetic field is polarized along the $y$-direction with a maximum amplitude of $280$~mT. The pump polarization was chosen to maximize the Zeeman torque on the sample magnetisation by the pump pulse. We characterized the THz pump pulse spectrum using electro-optical sampling with a 50 $\mu$m-thick GaP crystal. The broadband spectrum of the THz pump is centered at 2 THz with a bandwidth exceeding 1 THz (see Fig.~\ref{fig:THZ_pulse}). The THz pump and $800$~nm probe beams were collinearly incident on the sample at a 45-degree angle. {The pump-induced changes in the Kerr rotation angle ($\Delta$$\theta_K$) were monitored using a balanced detection setup.}

{At room temperature, magnetic contributions are typically isolated by recording time traces under static dc bias fields of equal magnitude and opposite polarity and taking their difference. This approach relies on an electromagnet to reverse the bias field, and the resulting difference signal selects components that are odd under bias field reversal. Implementing such a scheme inside a cryostat is technically challenging.
To overcome this limitation, we keep the bias field constant and reverse the polarity of the THz pump magnetic field. Two time traces are recorded, one for each THz polarity, and their difference isolates the response components that are odd with respect to the THz magnetic field, including the coherent magnetic precession and the concomitant magnetisation response~\cite{Unikandanunni_2022,hudl2019nonlinear}. Comparison at room-temperature confirms that the difference signal obtained by THz-polarity reversal closely matches the purely magnetic signal measured from reversing the dc bias field, indicating that the extracted difference signal is predominantly magnetic in origin. 

\begin{figure}[t]
    \centering
    \includegraphics[width=0.49\textwidth]{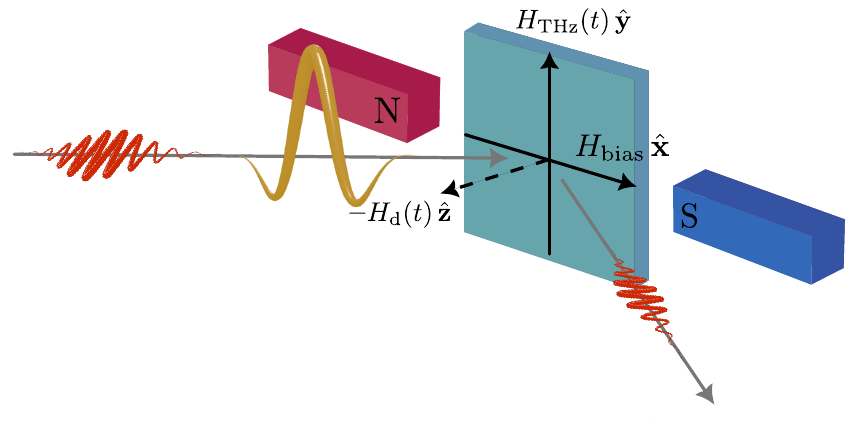}
    \caption{\textbf{Experimental schematic:} THz pump-Magneto-Optical Kerr Effect (MOKE) probe experiment on epitaxial Cobalt thin film. The THz pump and optical probe are collinear with an angle of incidence of 45 degrees. The directions of external bias magnetic field ($H_{\mathrm{bias}}\,\hat{\mathbf{x}}$), THz magnetic field ($H_{\mathrm{THz}}(t)\,\hat{\mathbf{y}}$), and the demagnetizing field ($-H_{\mathrm{d}}(t)\,\hat{\mathbf{z}}$) are shown.}
    \label{fig:experimental_setup}
\end{figure}

The sum of the two traces isolates the components that are even with respect to the THz field, which include non-magnetic phonon dynamics and other even-order magnetic contributions. The latter are negligible within our experimental time window. Reversal of the THz field polarity is achieved by rotating the THz generation crystal by 180$^{\circ}$. This introduces a temporal shift of 165 fs due to crystal-thickness inhomogeneity. The temporal shift is corrected before calculating difference and sum signals.

\medskip

\noindent {\bf Sample information}

\noindent 
The epitaxial cobalt sample under study has a face-centered cubic (fcc) structure with a thickness of $15$~nm. The sample stack, from top to bottom, consists of Pt (3 nm)/Co (15 nm)/Cu (20 nm)/MgO (substrate). The 3-nm-thick platinum layer serves as a cap layer, while the 20-nm-thick copper layer functions as a seed layer for the epitaxial growth on the MgO substrate.
Orientation-dependent static magneto-optical characterizations revealed that the sample exhibits minimal magnetocrystalline anisotropy {within the plane}. In our experiments, the external bias magnetic field was applied along the [100] crystalline direction of the fcc structure. The magnetic field required to saturate the sample magnetisation is $<100$~mT. Therefore, we applied a field of $100$~mT in our experiments to fully align the magnetisation along the $x$-direction. The static magneto-optical characterization of the sample is detailed in our previous work~\cite{Unikandanunni_2022}.

\begin{figure}[t]
    \centering
    \includegraphics[width=0.45\textwidth]{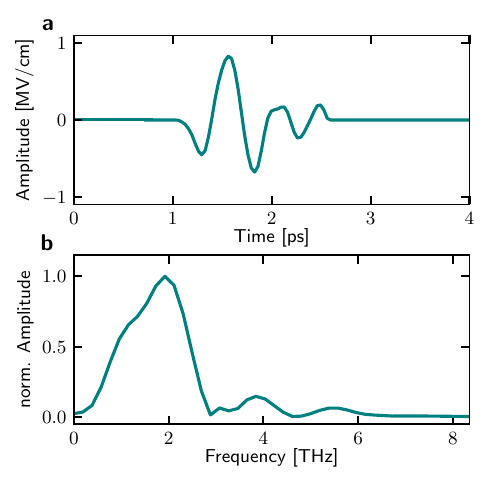}
    \caption{\textbf{Electro-optic sampling of the THz pump pulse:}
    {\textbf a.} Time-domain data: The time trace is zero-padded from 2.7 ps onward to eliminate internal reflections from the thin electro-optic crystal.
    {\textbf b.} Fast Fourier transform of the time-domain signal, revealing the pump spectrum.}
    \label{fig:THZ_pulse}
\end{figure}

\medskip

\noindent {\bf Technical simulation details}

\noindent For the plots in Fig.~\ref{fig:multipeak}\textbf{b}, 
all three equations, the LLG, iLLG and \mLLG~simulations are  run without stochastic noise, i.e. at zero temperature. 
The LLG and iLLG simulations are run with the same parameters (i.e. damping, inertial time, initial state) as the \mLLG~simulations to guarantee a fair comparison.
The effect of temperature on the observed THz-frequency spectrum for the \mLLG~equation is discussed in the main text (cf. Fig.~\ref{fig:temperaturedep}\textbf{b}) and further details are given in the~\supp. 

Simulating the iLLG equation follows the same procedure as previously outlined in Refs.~\cite{Neeraj_2020,Unikandanunni_2022}. 
The simulation of the \mLLG~equation is done via the open-source software \textsc{SpiDy}~\cite{Scali_2024}, which simulates a dimensionless version of Eq.~\eqref{eq:mLLG}.
To rerun the simulation, one needs to input the simulation time $\tilde{t} = \tilde{\omega}_{\mathrm{L}}t$ and the Lorentzian parameters, $\tilde{\alp} = 8\pi^3 \alp/\tilde{\omega}_{\mathrm{L}}^3 \simeq 1.1\cdot10^7$, $\tilde{\omega}_0 = 2\pi\omega_0/\tilde{\omega}_{\mathrm{L}} \simeq 151.7$, and $\tilde{\Gamma} = 2\pi\Gamma/\tilde{\omega}_{\mathrm{L}}\simeq 7.21$. These are scaled via an effective Larmor frequency  $\tilde{\omega}_{\mathrm{L}}$, which is defined for a fixed field strength $|\mathbf{H}| = 1.0$~T, thus, $\tilde{\omega}_{\mathrm{L}} = |\gamma|\cdot1\,\mathrm{T} \approx 1.76\cdot10^{11}$~rad/s, with $|\gamma| \approx 1.76\cdot10^{11}\,$rad/s/T being the electron gyromagnetic ratio.
This effective Larmor frequency $\tilde{\omega}_{\mathrm{L}}$ is solely used to transfer the dimensioned experimental parameters given in the main text into dimensionless simulation parameters, and vice versa.
The $\textit{real}$ Larmor frequency $\omega_{\mathrm{L}}$ of the simulated system is set by the effective magnetic field strength $\omega_{\mathrm{L}} = \gamma|\mathbf{H}_{\mathrm{eff}}|$, see Eq.~\eqref{eq:mLLG}.

The simulation incorporates the magnetic fields, where the effective  field can be written as the following sum  
\begin{equation}
    \mathbf{H}_{\mathrm{eff}}(t) = (H_{\mathrm{bias}} + H_{\mathrm{K}})\hat{\mathbf{x}} + H_{\mathrm{THz}}(t)\hat{\mathbf{y}}+H_{\mathrm{d}}(t)\hat{\mathbf{z}}.
\end{equation}
In the experiment, the external bias field $H_{\mathrm{bias}} = 100\,\mathrm{mT}$ is applied along the $x$-direction.
We neglect the anisotropy field $H_{\mathrm{K}}$, as it is small for the fcc Co sample compared to the other magnetic field components~\cite{Unikandanunni_2022}.
The demagnetisation field is given by~\cite{hudl2019nonlinear}
\begin{equation}
    \mathbf{H}_{\mathrm{d}}(t) = -N_{x,0} M_x(t) \hat{\mathbf{x}} - N_{y,0} M_y(t)\hat{\mathbf{y}} - N_{z,0} M_z(t)\hat{\mathbf{z}},
    \label{eq:biasfield}
\end{equation}
where $N_{x,0}$, $N_{y,0}$, and $N_{z,0}$ are the demagnetisation field factors. The normalized spin vector components are $\mathbf{M}(t) = (M_x(t),M_y(t),M_z(t))$.
Here, we have $N_{x,0} = N_{y,0} = 0$ and $N_{z,0} = 1.37$~T, therefore, Eq.~\eqref{eq:biasfield} reduces to $\mathbf{H}_{\mathrm{d}} = - N_{z,0}\,M_z(t)/M_0\,\hat{\mathbf{z}}$, where $M_0$ is the length of the macrospin, see Fig.~\ref{fig:experimental_setup}.
For a detailed numerical and experimental study of the above defined demagnetisation field, see Ref.~\cite{hudl2019nonlinear}.
Our non-equilibrium dynamics simulations start once the THz-pulse field has pushed the spin out of equilibrium. Thus, the THz field in the simulation is off, $H_{\mathrm{THz}}(t) = 0$, while we set the initial conditions to $\mathbf{M} (0) = (\sqrt{0.98},\sqrt{0.02},0.0)$ (cf. section on the initial state below).

We numerically integrate the \mLLG~equation, transforming the integro-differential equation~\eqref{eq:mLLG} into a system of first-order coupled differential equations~\cite{Anders_2022,Scali_2024},
\begin{align}
    \frac{\mathrm{d}\mathbf{M}(t)}{\mathrm{d}t} =&\, \mathbf{M}(t) \times \left(\mathbf{H}_{\mathrm{eff}}+\frac{1}{\sqrt{M_0}}\mathbf{h}_{\mathrm{th}}(t)+\mathbf{V}(t)\right), \nonumber \label{eq:numerics}\\
    \frac{\mathrm{d}\mathbf{V}(t)}{\mathrm{d}t} =&\,\mathbf{W}(t),
    \\
    \frac{\mathrm{d}\mathbf{W}(t)}{\mathrm{d}t} =&\,-\nu_0^2 \mathbf{V}(t) - \Gamma\mathbf{W}(t) + \alp \mathbf{M}(t), \nonumber
\end{align}
here $|\mathbf{M}(0)|$ is assumed to be normalized.
$\mathbf{V}(t)$ and $\mathbf{W}(t)$ are dummy vectors to incorporate the memory kernel $K(t-t')$. 
Mathematically, this step is known as Markovian embedding, where the full non-Markovian dynamics is solved via coupled Markovian equations~\cite{Wang_1945,Mori_1965}.
This introduces the need for nine initial conditions (three for $\mathbf{M}$ and three for $\mathbf{V}$ and $\mathbf{W}$), compared to three or six for the LLG or iLLG equation, respectively.
To simplify, we set $\mathbf{V}(0) = \mathbf{0}$ and $\mathbf{W}(0) = \mathbf{0}$, representing an initial state without memory effects or momentum in the auxiliary variables. 
We choose $\mathbf{M}(0)$ slightly out of equilibrium to model realistic dynamics, where the system is perturbed from perfect alignment, as typical for pump-probe experiments when the pump pulse has passed.

Similarly to the amplitude of the time-resolved MOKE signal, see above, the \textit{absolute} amplitude of the simulated THz-frequency spectrum is not directly informative. 
It depends, e.g., on the time-step choice and the exact initial conditions. 
For visualization purposes, we here chose the amplitudes of the experimental and simulation spectra such that their first THz-frequency peak amplitudes match, as seen in Fig.~\ref{fig:multipeak}\textbf{b}.

\medskip

\section*{Data availability}
\noindent The data supporting the findings of this study are available upon reasonable request from the corresponding authors.

\medskip

\section*{Code availability}
\noindent The code supporting the findings of this study is available upon reasonable request from the corresponding authors.

\bibliography{main.bib}

\medskip 

\begin{acknowledgments}
\noindent We thank Federico Cerisola for assistance with using the SpiDy code (see \supp).
JA thanks Simon Horsley for anticipating a connection between  \cite{Neeraj_2020} and \cite{Anders_2022}. 
We thank Fried-Conrad Weber and Tobias Kampfrath for discussions on the topic. 
We thank Brendon Lovett for  comments on an early draft of the manuscript.
S.B. acknowledges support from the Italian Ministry of University and Research. This study was carried out within the PRIN 2022 cod. 2022ZRLA8F, and received funding from the European Union Next-GenerationEU - National Recovery and Resilience Plan (NRRP) – MISSION 4 COMPONENT 2, INVESTIMENT 1.1 Fondo per il Programma Nazionale di Ricerca e Progetti di Rilevante Interesse Nazionale (PRIN) – CUP: H53D23000890006.
VU acknowledges the support of the SNSF Sinergia project (CRSII5 213533).
MB acknowledges funding by the Deutsche Forschungsgemeinschaft (DFG, German Research Foundation) – TRR 227 – Project-No. 328545488, project A10.
JA acknowledges support from EPSRC grant EP/R045577/1, UKRI grant UKRI3778, and the Royal Society. 
FH and JA acknowledge support from the DFG, grants 513075417 and 384846402.
JA and MB acknowledge funding by the DFG, grant 510943930 (CRC1636).
\end{acknowledgments}

\medskip 

\section*{Author Contributions}
\noindent J.A. conceived the initial idea, which was further developed by F.H., V.U., S.B. and J.A.
F.H. performed and analyzed the simulations. 
S.B. and V.U. designed and conducted the experiments, and analysed the data. 
E.F. fabricated the samples. 
F.H., V.U. and J.A. wrote the first draft of the manuscript. 
M.B. contributed key points to the interpretation of the observations. 
All authors discussed the results and commented on the manuscript. 
S.B. and J.A. supervised the project.

\medskip 

\section*{Competing Interests}
\noindent The authors declare no competing interests.


\cleardoublepage



    





\section*{Appendix}

\medskip

\appendix
\section{Amplitude of the time-resolved MOKE signal}
\label{supp:amplitude_of_MOKE_signal}
The time-resolved MOKE signal represents the magneto-optical response, which is proportional to the magnetisation dynamics induced by the pump pulse. The relationship between the time-dependent Kerr angle  $\Delta \theta_K$ and the change in magnetisation $\Delta M$ in the weak-perturbation limit is given by
\begin{equation}
    \Delta \theta_K(t) = \theta_{K0}  \frac{\Delta M(t)}{M_0},
    \label{eq:delta_theta_k}
\end{equation}
where $\theta_{K0}$ is the static magneto-optic Kerr response and $M_0$ is the total magnetisation of the sample~\cite{unikandanunni2021ultrafast}. The static Kerr response $\theta_{K0}$ itself depends on various probe parameters, including but not limited to the probe wavelength, angle of incidence, polarization, and the sample's refractive index at the probe wavelength \cite{mansuripur1990measuring}. Consequently, the absolute magnitude of the experimentally measured magneto-optical response is not intrinsically significant and can vary depending on the probing scheme. 

\begin{figure}[b]
    \centering
    \includegraphics[width=0.48\textwidth]{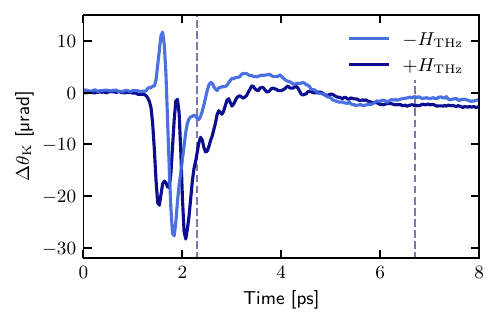}
    \caption{\textbf{Signal for positive and negative polarized pump pulse:} Kerr signal for positive $+\mu_0H_{\mathrm{THz}}\hat{\mathbf{y}}$ (dark blue solid line) and negative $-\mu_0H_{\mathrm{THz}}\hat{\mathbf{y}}$ (light blue solid line) polarized pump pulse, respectively.}
    \label{fig:positive_and_negative_signal}
\end{figure}

\section{Data analysis}
\label{supp:data_analysis}

To perform the Fourier transform, we used time-domain data from a time delay of 2.3 ps to 6.7 ps, where the THz-driven coherent precession has essentially vanished, and the MOKE signal oscillations were clearly visible (see \EDF~\ref{fig:sup3}). The data was then multiplied by a Blackman window centered at 4.5 ps with a width of 4.4 ps to minimize spectral leakage. Finally, a Fast Fourier Transform (FFT) was applied to the resulting time-domain signal to obtain the frequency spectrum.

The error bars in Fig.~\ref{fig:multipeak} reflect the statistical instrument noise and are given by $2\sigma_{\mathrm{FFT}}$, where the standard deviation in Fourier space is calculated as 
\begin{equation}
    \sigma_{\mathrm{FFT}} = C\sigma_t \sqrt{\sum_{n=0}^{N-1}w_n^2}.
\end{equation}
Here $\sigma_t$ is the standard deviation of the lock-in-integrated time-domain traces in the pre-pump region, $C$ is the FFT normalization, and $w_n$ is the windowing function as detailed below.

The static MOKE signal only varies by approximately 1\% over the investigated temperature range (Fig.~\ref{fig:temperaturedep}), and the detector noise and laser fluctuations are not expected to depend significantly on the sample temperature, hence, the error bars are assumed to be temperature independent.

Rotating the THz polarization by 180$^\circ$ will produce a small error, which is not included in the error bars in Fig.~\ref{fig:multipeak}.
The THz field polarization is flipped by a rotation of the inhomogeneous organic crystal. This means the experiment works at a slightly different spot in the crystal, which leads to a few degrees change in THz polarization and in the THz generation efficiency.
This will result in small leakage between  the difference and the sum signal.

\section{Sum (non-magnetic) signal and phonon details}
\label{supp:nonmagnetic_phonons}

\begin{figure}
    \centering
    \includegraphics[width=0.48\textwidth]{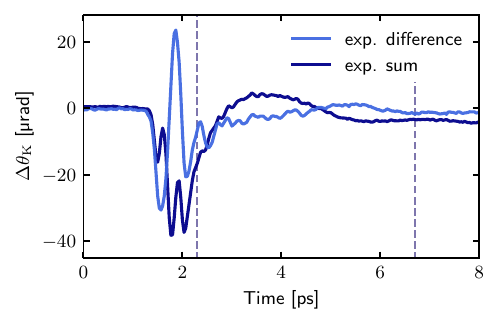}
    \caption{\textbf{Sum and difference time traces:} Time trace obtained by subtracting (magnetic, light blue) and summing (non-magnetic, dark blue) signals, obtained for positive and negative THz pump polarities, respectively, at a temperature of $20$~K. The Fourier transforms of the two time traces, for the time window $2.3- 6.7$~ps (dashed lines), are presented in main text Fig.~\ref{fig:multipeak}{\textbf a} and {\textbf b}. 
    }
    \label{fig:sup3}
\end{figure}

{As discussed in the main text, we refer to the sum and difference signal of the two THz polarities as non-magnetic and magnetic signal, respectively.} 
The experimentally determined {non-magnetic} and {magnetic} contributions at 20 K are shown in \EDF~\ref{fig:sup3}. The Fourier transform of the {non-magnetic} signal, presented in Fig.~\ref{fig:multipeak}{\textbf a}, reveals a clear excitation peak at a frequency of 4.2 THz and a weaker excitation peak at 5.7~THz. Here we discuss the physical origin of this signal.

The phonon DOS of fcc Co shows three dominant peaks at approximately 4.2 THz, 5.7 THz, and 8 THz, which are associated with two zone-edge transverse and one longitudinal acoustic branches, respectively~\cite{Lizrraga_2017,Carva_2013}.
Recent energy- and momentum-resolved studies have demonstrated the critical role of zone-edge phonons in ultrafast spin–lattice interactions in metallic ferromagnets~\cite{maldonado2020tracking,Shokeen_2024}. In particular, transverse acoustic phonons have been identified as efficient angular momentum sinks on sub-picosecond timescales~\cite{Dornes_2019}, while longitudinal acoustic phonons, though capable of transporting energy, cannot carry angular momentum and therefore do not contribute significantly to spin–lattice angular momentum transfer. 
These microscopic mechanisms underscore the unique importance of $4.2$~THz transverse acoustic phonons and indicate their active participation in  ultrafast spin dynamics.

\begin{figure}
    \centering
    \includegraphics[width=0.48\textwidth]{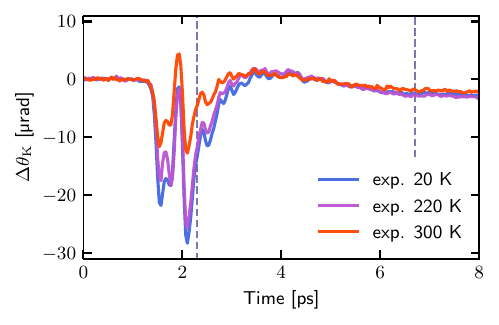}
    \caption{\textbf{Time traces of temperature-dependent magnetisation dynamics:}  
    This is for a fixed polarity of THz field, and thus the trace contains magnetic and non-magnetic contributions.  In addition to coherent precession (slow oscillation) and THz-frequencies (fast oscillations), a ``non-magnetic'' response (deep dips around 2 ps) are observed. Spectra shown in main text Fig.~\ref{fig:temperaturedep} are obtained via FFT of the time trace window $2.3 - 6.7$~ps.} 
    \label{fig:sup2}
\end{figure}

\section{Experiments on additional samples}
\label{supp:additional_samples}

\begin{figure}
    \centering
    \includegraphics[width=0.48\textwidth]{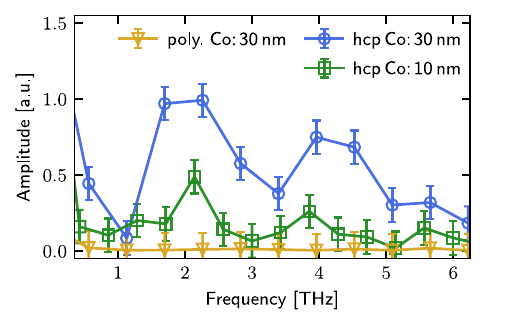}
    \caption{\textbf{FFT of THz pump-MOKE probe measurements of three Co samples:} 30-nm hcp-Co (blue circles), 10-nm hcp-Co (green squares), and 30-nm polycrystalline Co (yellow triangles). The epitaxial hcp-Co samples exhibit distinct THz spectral peaks in the FFT, while the polycrystalline film shows no resolvable peaks. All three measurements are performed at room temperature. Error bars indicate $2\sigma$ statistical error.}
    \label{fig:polycrystalline}
\end{figure}
{To confirm that the observed dynamics arise from the intrinsic magnetic response of the sample - rather than from spectral structure in the THz pump - we performed complementary measurements on three additional samples using the identical experimental setup. These measurements were carried out at room temperature. Therefore, unlike the low-temperature experiments described in the main text, we isolated the magnetic contribution to the signal by inverting the external bias field instead of the THz polarization.
The experimental spectra of all three control samples are shown together in the \EDF~\ref{fig:polycrystalline}. The first control sample was a 30-nm-thick epitaxially grown hcp-Co film (blue circles), which exhibits three clear THz spectral peaks at 2.1 THz, 3.9 THz, and 5.6 THz, consistent with previous reports~\cite{Unikandanunni_2022}. 
The second sample was a 10-nm-thick epitaxial hcp-Co film (green squares). It shows spectral peaks at approximately the same frequencies, but with amplitudes reduced by roughly a factor of two. This reduction does not scale linearly with thickness because the THz penetration depth in Co ($\sim$ 20 nm) lies between the two sample thicknesses: i.e. the 10 nm film is fully excited, whereas only approximately two-thirds of the 30 nm film participate in the dynamics. Despite the amplitude difference, both epitaxial hcp-Co samples exhibit qualitatively similar behavior that is nonetheless clearly distinct from the response of the epitaxial fcc-Co sample discussed in the main text.
The third sample was a 30-nm-thick polycrystalline Co film (yellow triangles). Here, no THz-frequency spectral peaks were detected. 
This stark contrast with the epitaxial samples highlights that high crystalline quality is essential for observing a coherent, non-Markovian magnetisation dynamics.
Together, these control measurements demonstrate that the dynamics reported in the main text originate from the sample’s intrinsic magnetisation response. The measured behavior depends sensitively on crystal structure~\cite{Li_2012} and disappears when the crystalline order is reduced.}

\section{Parameter variation}

\begin{figure}
    \centering
    \includegraphics[width=0.48\textwidth]{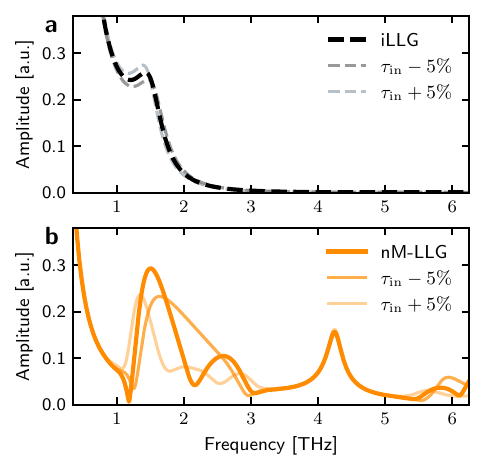}
    \caption{\textbf{Parameter varied spectrum.}  
    Thick lines show spectra identical to Fig.~\ref{fig:multipeak}\textbf{b} while the more transparent lines show changes of the spectra when the inertial time $\tau_{\mathrm{in}}$ is varied by $\pm 5\%$. In \textbf{a}. the simulation of the iLLG equation and in \textbf{b}. the simulation of  the \mLLG~equation. 
    }
    \label{fig:parameter_variation}
\end{figure}

We here compare the robustness of the obtained frequency spectra for the iLLG and \mLLG~equation for changes in the inertial time $\tau_{\mathrm{in}} = 0.8$~ps. 
For changes of $\pm 5\%$ the spectra are shown in \EDF~\ref{fig:parameter_variation}. 
We observe that the frequency of the \textit{single} inertial peak in the spectrum of the iLLG equation is inversely proportional with $\tau_{\mathrm{in}}$, as previously reported e.g. in Ref.~\cite{Olive_2015}.
Likewise we find that the lowest frequency peak in the THz range in the spectrum of the \mLLG~equation changes; however, the change is more pronounced and other parts of the spectrum change as well.
The frequency peak at $4.2$~THz is very robust. This arises because the non-magnetic $\nu_0$ enters as a parameter in the  memory kernel, see Eq.~\eqref{eq:nM-LLG_kernel}, which hence oscillates with $\nu_1 = \sqrt{\nu_0^2 - {\Gamma^2/ 4}} \approx \nu_0$, for $\Gamma \ll \nu_0$.  
A movie, showing the change of the THz-frequency spectrum arising from the \mLLG-equation when it is solved for different inertial times $\tau_{\mathrm{in}}$ while holding $\alpha$ and $\nu_0$ fixed ($\alp$ and $\Gamma$ are changed accordingly) is available at~\cite{movie_2025}.
We note that the specific THz-frequency spectrum is quite sensitive to variation, both experimentally \cite{Neeraj_2020} and theoretically, see \EDF~\ref{fig:parameter_variation}. This points at the reasons of why it has been rather difficult for the community to measure THz-frequency spectra. 
Despite the parameter-sensitivity of the detailed spectrum, we highlight that the theory consistently predicts, for a wide range of  Lorentzian parameters, a structured multi-peak spectral THz-frequency response. This indicates the possibility to characterize the THz-frequency spectrum of other ferromagnetic materials in the future.

\begin{figure}
   \centering
\includegraphics[width=0.48\textwidth]{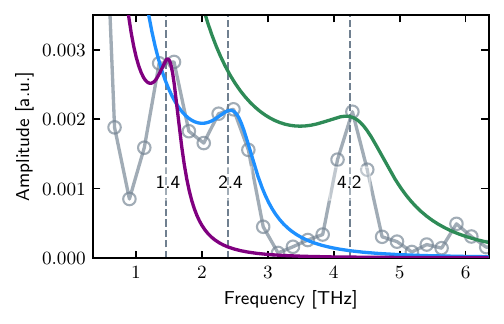}
   \caption{\textbf{Magnetic response in the inertial regime for three  values of inertia.} Fourier spectrum of experimental magnetic signal (grey) and Fourier spectra of simulated inertial LLG equation with fixed Gilbert damping $\alpha = 0.15$ and three values of $\tau_{\rm in} $  (three coloured lines). The value $\tau_{\rm in} =0.8$~ps (purple) is used for the plot in main text Fig.~\ref{fig:multipeak}\textbf{b}. Inertial dynamics with this value predicts a spectral peak at 1.4~THz, which matches the first experimental peak (grey), but does not produce any of the remaining experimentally observed peaks at 2.4~THz and 4.2~THz.
    Decreasing the inertial timescale $\tau_{\rm in}$ in the simulation to either $0.43$~ps (blue) or $0.245$~ps (green), results in the single Fourier peak to shift to either 2.4~THz or to 4.2~THz, respectively. 
    All three inertial spectra (coloured) show a single peak and do not produce the complex spectrum observed experimentally.
   }
   \label{fig:mutiple_inertial_sim}
\end{figure}

\section{Form of memory kernels in time}
\label{supp:memory_kernels}

\begin{figure}
   \centering
\includegraphics[width=0.48\textwidth]{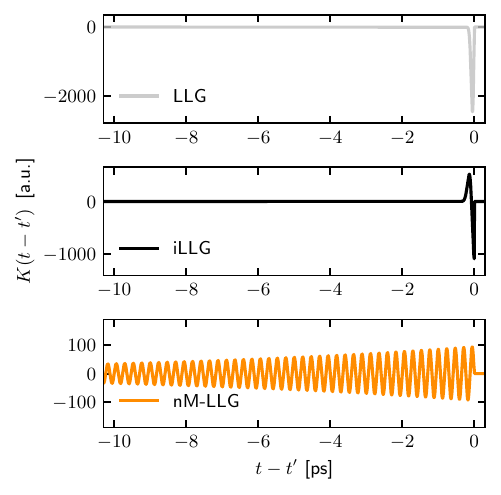}
   \caption{\textbf{Memory kernels in time:}
   Plot of memory kernels $K(t-t')$, which when inserted into Eq.\eqref{eq:mLLG}, result in the three equations, LLG, iLLG, nM-LLG, respectively. The kernels shown here lead to the Fourier spectra plotted in Fig.~\ref{fig:multipeak}{\textbf b}. 
   The standard LLG equation is time-local and  Markovian. 
   The kernel for the \mLLG~equation has an extended memory time of $\sim 10$~ps, which leads to non-Markovian dynamics on this timescale. 
   }
   \label{fig:kernels}
\end{figure}

In this manuscript we discussed that the \mLLG~equation with its memory kernel is capable to capture the experimentally observed complex frequency spectrum. 
The kernel of the Lorentzian spectral density is given by
\begin{equation}
    K(\tau) = \Theta(\tau)\alp e^{-\Gamma\tau/2}\frac{\sin(\nu_1\tau)}{\nu_1},
    \label{eq:nM-LLG_kernel}
\end{equation}
where $\tau = t-t'$ and $\nu_1 = \sqrt{\nu_0^2-\Gamma^2/4}$. For the above discussed Lorentzian parameters the kernel is plotted in the bottom panel of \EDF~\ref{fig:kernels}.

In Ref.~\cite{Anders_2022} the LLG-limit (time-local, Markovian) of Eq.~\eqref{eq:mLLG} is discussed. To obtain the desired limit, one chooses a spectral density which is linear in frequency, i.e. $I(\nu) \propto \alpha\nu$, where $\alpha$ is the damping parameter. 
This leads to the following memory kernel
\begin{equation}
    K(\tau) = -\alpha\Theta(\tau)\partial_{\tau}\delta(\tau-\epsilon_{+}),
    \label{eq:LLG_kernel}
\end{equation}
where $\epsilon_+$ is an infinitesimal positive constant which is set to zero at the calculation's end.

Following a similar logic one can construct a memory kernel which leads to the  iLLG equation
\begin{equation}
    K(\tau) = \alpha\Theta(\tau)\left(\partial_\tau\delta(\tau-\epsilon_+) +\tau_{\mathrm{in}}\partial_\tau^2\delta(\tau-\epsilon_+) \right).
    \label{eq:iLLG_kernel}
\end{equation}
The memory kernel for the LLG and iLLG equation are shown in the top and middle panel of \EDF~\ref{fig:kernels}, respectively. 
For illustrative purposes, the delta functions are approximated by a Gaussian.

\section{Multiple THz-frequencies predicted by susceptibility expansion}
\label{supp:susceptibility_expansion}

In the main text we gave a heuristic explanation for why the \mLLG~equation~\eqref{eq:mLLG} produces multiple THz-frequency peaks. 
Here we provide a more detailed educational description, following the approach proposed in~\cite{Cherkasskii_2020}.

The magnetic response arises from the memory-kernel $K(t-t')$ in Eq.~\eqref{eq:mLLG}, which is Taylor expanded in Eq.~\eqref{eq:mLLGexpanded}. 
The effective field $\mathbf{H}_{\mathrm{eff}}(t)$ is specified as in Eq.~\eqref{eq:mLLG}, however, the demagnetisation field is neglected for simplicity and we drop the $\mathrm{eff}$ index.
We wish to find the magnetic response in Fourier picture (indicated by $\tilde{\,}$); 
more specifically we wish to find the  susceptibility $\chi_j (\nu)$ which links the field to its response via
\begin{equation}
\tilde{M}_j  (\nu) = M_0 \, \chi_j  (\nu) \, \tilde{H}_j (\nu),
\end{equation}
for each spatial component $j$. 
We follow~\cite{Cherkasskii_2020} and linearize Eq.~\eqref{eq:mLLGexpanded} by neglecting higher-order products of $\mathbf{M}$ and $\mathbf{H}$. Applying the Fourier transform one obtains 
\begin{align}
    i\nu \, \mathbf{\tilde{M}}(\nu) =
    & -\nu_{\mathrm{L}} \, \hat{\mathbf{x}}\times \mathbf{\tilde{M}}(\nu) + \hat{\mathbf{x}}\times\mathbf{\tilde{H}}(\nu) \nonumber + \hat{\mathbf{x}}\times \sum_{m=1}^{\infty}\kappa_m \, (i\nu)^m \, \mathbf{\tilde{M}}(\nu),
\end{align}
with $\nu_{\mathrm{L}} = \omega_{\mathrm{L}}/(2\pi) = 2.8\,\mathrm{GHz}$ being the Larmor frequency for the external bias field of $H_{\mathrm{bias}} = 100$~mT.

Using $\tilde{M}_{\pm} = \tilde{M}_z \pm i \tilde{M}_y$ and $\tilde{H}_{\pm} = \tilde{H}_z \pm i \tilde{H}_y$, one arrives at the susceptibility $\chi_{\pm}(\omega)$ for positive $(+)$ and negative $(-)$ polarization directions,
\begin{equation}
    \chi_{\pm}(\nu) = \left({\nu_{\mathrm{L}} - \sum_{m=1}^{\infty}\kappa_m (i\nu)^m \pm \nu} \right)^{-1}.
    \label{eq:susceptibility}
\end{equation}
This susceptibility $\chi_{\pm}(\nu)$ diverges whenever the $m$-th order polynomial in Eq.~\eqref{eq:susceptibility} becomes zero. 
The absolute values of the real part of these roots give the frequency at which maxima (and also minima) are expected in the Fourier spectrum of the spin dynamics described by Eq.~\eqref{eq:mLLGexpanded}.

For the iLLG equation, the damping $\alpha$ and inertial time $\tau_{\mathrm{in}}$, relate to the two expansion coefficients $\kappa_{1,2}$ in Eq.~\eqref{eq:mLLGexpanded} as $\kappa_{1}/M_0 = -\alpha$ and $\kappa_{2}/M_0  = -\alpha \, \tau_\mathrm{in}$. 
For the parameter values specified above, one analytically finds roots at the precession frequency at $2.8$~GHz, and a single spectral THz-frequency at $1.4$~THz, consistent with the simulation shown in Fig.~\ref{fig:multipeak}{\textbf b}.
It is well-known, that the iLLG equation can only show a \textit{single} inertial peak, and its  frequency has previously been derived Refs.~\cite{Olive_2015,Cherkasskii_2020}.

In~\EDF~\ref{fig:mutiple_inertial_sim} we numerically find the magnetic response for three inertial dynamics. Each has a different value of the inertial timescale $\tau_{\rm in} =$ 0.8~ps, 0.43~ps and 0.245~ps, while keeping the Gilbert damping $\alpha$ fixed as before. For each $\tau_{\rm in}$, the Fourier spectrum of the dynamics is obtained and plotted (three coloured lines). 
The figure shows that the value of $\tau_{\rm in}$ does not increase the number of peaks - it is always a single peak. Instead, $\tau_{\rm in}$ determines \textit{where} the single peak occurs, i.e. its frequency.
The three values of $\tau_{\rm in}$ are fixed in a physically reasonable range -- such that each of the three peaks observed in the experimentally measured response (grey) is reproduced. 
The observed one-to-one correspondence between decreasing $\tau_{\rm in}$ and increasing single peak frequency is fully consistent with the established literature, where analytical derivations of a single inertial frequency as a function of $\tau_{\rm in}$ are provided~\cite{Olive_2015,Cherkasskii_2020}.

\medskip 

Now, the memory kernel $K(t-t')$ in the \mLLG~equation naturally leads to the presence of multiple expansion orders in Eq.~\eqref{eq:mLLGexpanded}. This implies higher order polynomials in Eq.~\eqref{eq:susceptibility}, which have a larger number of zeroes. Hence with a memory kernel present in the dynamics, we would analytically expect multiple spectral THz-frequency peaks. Specifically, for our Lorentzian memory kernel, the expansion coefficients $\kappa_{m}/M_0$ in Eq.~\eqref{eq:mLLGexpanded} are \cite{Anders_2022}  $\kappa_{m}/M_0= (-1)^m\alp/\left(\nu_1\nu_0^{2(m+1)}\right) \, \mathrm{Im}[((\Gamma/2)+i\nu_1)^{m+1}]$, where $\nu_1 = \sqrt{\nu_0^2-\Gamma^2/4}$. 
While this series never breaks up, the coefficients decay with increasing order. Here we take into account the first six expansion orders. Solving for the zeroes of the sixth order polynomial in Eq.~\eqref{eq:susceptibility}, we find zeroes at $2.8$~GHz and $1.23$~THz -- which are close to the zeroes of the iLLG equation for the precession and the inertial peak, respectively. In addition, we find a (semi-degenerate) frequency at $2.45$~THz (and $2.46$~THz), which very well matches the second spectral THz-frequency peak found numerically in the spin dynamics, see Fig.~\ref{fig:multipeak}\textbf{b}. 
A further (semi-degenerate) zero lies at $1.82$~THz (and $1.85$~THz), which appears to correspond to a ``valley'' in Fig.~\ref{fig:multipeak}\textbf{b}.

In addition, one can analytically anticipate a further peak by looking at the \mLLG~equation and by identifying the oscillation frequency of the kernel $K(t-t')$, which carries over into the
spin dynamics. 
For a spectral density with a  Lorentzian peak at frequency $\nu_0$ and decay rate $\Gamma$, the kernel oscillation frequency is given by $\nu_1 = \sqrt{\nu^2_0-\Gamma^2/4}$, see Eq.~\eqref{eq:nM-LLG_kernel} in the previous section. The resulting numerical value is 4.2~THz, which matches the third experimentally measured peak.

We note that while this description is educational, it does not reliably predict the exact frequencies of the higher-order peaks.
This is because the analytical approach is based on several assumptions, including a linearization of Eq.~\eqref{eq:mLLGexpanded} and a time-local expansion of the memory kernel.
Further untangling the mathematical interplay of non-linear effects together with the memory kernel will be the subject for future research.

\medskip

\section{Temperature-dependent simulations}
\label{supp:demag_field_temperature}
Adding a demagnetisation field $\mathbf{H}_\mathrm{d} (t) = - N_{z,0} \, {M_z(t)/ M_0} \, \hat{\mathbf{z}}$ to the bias field $\mathbf{H}_{\rm bias} = \hat{\mathbf{x}}\cdot0.1~\mbox{T}$ is a way of introducing the geometry of the film, with $N_{z,0}$ being the demagnetisation field factor at zero Kelvin.
The effect of $\mathbf{H}_{\mathrm{d}}(t)$ is to counteract any magnetisation pointing outside the film, i.e. $\mathbf{H}_\mathrm{d}(t)$ acts whenever there is a non-zero $M_z$ component. 
The demagnetisation field factor $N_{z,0}$ is taken for temperature of 0 Kelvin as $N_{z,0} = 1.37$~T. With increasing temperature, the magnitude of the magnetisation, $\langle M_x\rangle_\mathrm{eq}(T)$, decreases and the counteracting demagnetisation field is then proportionally reduced~\cite{Evans_2014}, i.e. here $\mathbf{H}_\mathrm{d} (t, T) = - N_T  \, {M_z(t) /M_0} \,  \hat{\mathbf{z}}$ with $N_{\mathrm{T}}=N_{z,0} \, {\langle M_x\rangle_\mathrm{eq} (T) \over \langle M_x\rangle_\mathrm{eq}(0)}$.
Here $\langle M_x\rangle_\mathrm{eq} (T) $ is obtained by solving the dynamics to steady state, where $\langle M_y\rangle_{\mathrm{eq}} = \langle M_z\rangle_{\mathrm{eq}}=0$, including the demagnetisation field of $\mathbf{H}_{\mathrm{d}}(t,T=0K)$. Second order corrections could be taken into account self-consistently; however, these corrections are negligible here. 
This yields the following demagnetisation field factors for the three simulation temperatures in the main text: $N_{\mathrm{T}} (\tilde{T} = 20) \approx 0.98\,N_{z,0}$, $N_{\mathrm{T}} (\tilde{T} = 220) \approx 0.8\,N_{z,0}$, and $N_{\mathrm{T}} (\tilde{T} = 300) \approx 0.73\,N_{z,0}$.


The \mLLG~simulation is carried out for a single magnetic moment $M_0$. This can represent either a single spin or, in the macrospin approximation, an effective moment associated with an ensemble of interacting spins. Due to the \mLLG~equation only depending on the combined scale $M_0 T_{\sf sim}$~\cite{Anders_2022}, the simulation temperature  $T_{\sf sim}$ corresponding to $T_{\sf exp}$ is not specified without knowing the magnetic moment in the macrospin approximation. Knowing this analytically is not feasible -- it would require specifying the number of spins, their individual moments and their interactions. Instead of doing so, we set the magnetic moment as $M_0 = 1 \hbar$ and chose the simulation temperature proportionally to the experimental temperature $T_{\sf sim} = T_{\sf exp} \, c/\mbox{K}$, with $c$ a proportionality constant.  We choose $c =1\cdot10^4 \,\hbar\tilde{\omega}_{\mathrm{L}}/k_{\mathrm{B}}$. 
This corresponds to a macrospin length of  $M_{\sf mac}  \approx 1.3 \cdot  10^4 \,  \hbar$.

\setcounter{figure}{0}
\renewcommand{\figurename}{Supplementary Figure}
\renewcommand{\thefigure}{S\arabic{figure}}

\end{document}